\documentclass[11pt,italian,a4widepaper]{article}

\usepackage{graphics,graphicx}
\usepackage{amssymb}
\usepackage{amsmath}
\usepackage{graphicx}
\usepackage{enumerate}
\usepackage{tabularx}
\usepackage{multirow}
\usepackage{subfigure}

\usepackage[dvips]{color}

\begin{document}

\newcommand{\singlespace}{\baselineskip=12pt
\lineskiplimit=0pt \lineskip=0pt }
\def\ds{\displaystyle}

\newcommand{\beq}{\begin{equation}}
\newcommand{\eeq}{\end{equation}}
\newcommand{\lb}{\label}
\newcommand{\beqar}{\begin{eqnarray}}
\newcommand{\eeqar}{\end{eqnarray}}
\newcommand{\und}{\underline}
\newcommand{\diam}{\stackrel{\scriptscriptstyle \diamond}}

\newcommand{\Ehat}{\hat{E}}
\newcommand{\Ahat}{\hat{A}}
\newcommand{\khat}{\hat{k}}
\newcommand{\muhat}{\hat{\mu}}
\newcommand{\mc}{M^{\scriptscriptstyle C}}
\newcommand{\mt}{M^{\scriptscriptstyle T}}
\newcommand{\mei}{M^{\scriptscriptstyle M,EI}}
\newcommand{\mec}{M^{\scriptscriptstyle M,EC}}
\newcommand{\mw}{M^{\scriptscriptstyle W}}

\newenvironment{sistema}%
{\left\lbrace\begin{array}{@{}l@{}}}%
{\end{array}\right.}

\def\ob{{\, \underline{\otimes} \,}}
\def\scalp{\mbox{\boldmath$\, \cdot \,$}}
\def\gdp{\makebox{\raisebox{-.215ex}{$\Box$}\hspace{-.778em}$\times$}}
\def\bob{\makebox{\raisebox{-.215ex}{$\Box$}\hspace{-.73em}$\scalp$}}

\def\c{{\circ}}

\def\bA{\mbox{\boldmath${\it A}$}}
\def\ba{\mbox{\boldmath${\it a}$}}
\def\bB{\mbox{\boldmath${\it B}$}}
\def\bb{\mbox{\boldmath${\it b}$}}
\def\bC{\mbox{\boldmath${\it C}$}}
\def\bc{\mbox{\boldmath${\it c}$}}
\def\bD{\mbox{\boldmath${\it D}$}}
\def\bd{\mbox{\boldmath${\it d}$}}
\def\bE{\mbox{\boldmath${\it E}$}}
\def\be{\mbox{\boldmath${\it e}$}}
\def\bF{\mbox{\boldmath${\it F}$}}
\def\bff{\mbox{\boldmath${\it f}$}}
\def\bG{\mbox{\boldmath${\it G}$}}
\def\bg{\mbox{\boldmath${\it g}$}}
\def\bH{\mbox{\boldmath${\it H}$}}
\def\bh{\mbox{\boldmath${\it h}$}}
\def\bi{\mbox{\boldmath${\it i}$}}
\def\bI{\mbox{\boldmath${\it I}$}}
\def\bj{\mbox{\boldmath${\it j}$}}
\def\bK{\mbox{\boldmath${\it K}$}}
\def\bk{\mbox{\boldmath${\it k}$}}
\def\bL{\mbox{\boldmath${\it L}$}}
\def\bl{\mbox{\boldmath${\it l}$}}
\def\bM{\mbox{\boldmath${\it M}$}}
\def\bm{\mbox{\boldmath${\it m}$}}
\def\bN{\mbox{\boldmath${\it N}$}}
\def\bn{\mbox{\boldmath${\it n}$}}
\def\b0{\mbox{\boldmath${0}$}}
\def\bo{\mbox{\boldmath${\it o}$}}
\def\bP{\mbox{\boldmath${\it P}$}}
\def\bp{\mbox{\boldmath${\it p}$}}
\def\bQ{\mbox{\boldmath${\it Q}$}}
\def\bq{\mbox{\boldmath${\it q}$}}
\def\br{\mbox{\boldmath${\it r}$}}
\def\bR{\mbox{\boldmath${\it R}$}}
\def\bS{\mbox{\boldmath${\it S}$}}
\def\bs{\mbox{\boldmath${\it s}$}}
\def\bT{\mbox{\boldmath${\it T}$}}
\def\bt{\mbox{\boldmath${\it t}$}}
\def\bU{\mbox{\boldmath${\it U}$}}
\def\bu{\mbox{\boldmath${\it u}$}}
\def\bv{\mbox{\boldmath${\it v}$}}
\def\bV{\mbox{\boldmath${\it V}$}}
\def\bw{\mbox{\boldmath${\it w}$}}
\def\bW{\mbox{\boldmath${\it W}$}}
\def\by{\mbox{\boldmath${\it y}$}}
\def\bX{\mbox{\boldmath${\it X}$}}
\def\bx{\mbox{\boldmath${\it x}$}}

\def\bbD{\overline{\bD}}
\def\bbL{\overline{\bL}}
\def\bbW{\overline{\bW}}

\def\bbeta{\mbox{\boldmath${\beta}$}}
\def\bepsilon{\mbox{\boldmath${\epsilon}$}}
\def\bvarepsilon{\mbox{\boldmath${\varepsilon}$}}
\def\bsigma{\mbox{\boldmath${\sigma}$}}
\def\bphi{\mbox{\boldmath${w}$}}
\def\bzeta{\mbox{\boldmath${\zeta}$}}

\def\bQ{\mbox{\boldmath $Q$}}

\def\Id{\mbox{\boldmath${\it I}$}}
\def\balpha{\mbox{\boldmath${\alpha}$}}
\def\bbeta{\mbox{\boldmath${\beta}$}}
\def\bGamma{\mbox{\boldmath${\Gamma}$}}
\def\bDelta{\mbox{\boldmath${\Delta}$}}
\def\bkappa{\mbox{\boldmath $\kappa$}}
\def\btau{\mbox{\boldmath $\tau$}}
\def\bnu{\mbox{\boldmath $\nu$}}
\def\bchi{\mbox{\boldmath${\chi}$}}
\def\bxi{\mbox{\boldmath${ \xi}$}}
\def\bXi{\mbox{\boldmath${\it  \Xi}$}}
\def\bsigma{\mbox{\boldmath${\sigma}$}}
\def\bvarsigma{\mbox{\boldmath${\varsigma}$}}
\def\bSigma{\mbox{\boldmath${\Sigma}$}}
\def\bupsilon{\mbox{\boldmath $\upsilon$}}
\def\bgamma{\mbox{\boldmath $\gamma$}}
\def\bTheta{\mbox{\boldmath $\Theta$}}

\def\tr{{\sf tr}}
\def\dev{{\sf dev}}
\def\div{{\sf div}}
\def\Div{{\sf Div}}
\def\Grad{{\sf Grad}}
\def\grad{{\sf grad}}
\def\Lin{{\sf Lin}}
\def\Orth{{\sf Orth}}
\def\Unim{{\sf Unim}}
\def\Sym{{\sf Sym}}

\def\msm{\mbox{${\mathsf m}$}}

\def\msM{\mbox{${\mathsf M}$}}
\def\msS{\mbox{${\mathsf S}$}}

\def\forA{\mathbb A}
\def\forB{\mathbb B}
\def\forC{\mathbb C}
\def\forE{\mathbb E}
\def\forL{\mathbb L}
\def\forN{\mathbb N}
\def\forR{\mathbb R}

\def\capA{\mbox{\boldmath${\mathsf A}$}}
\def\capB{\mbox{\boldmath${\mathsf B}$}}
\def\capC{\mbox{\boldmath${\mathsf C}$}}
\def\capD{\mbox{\boldmath${\mathsf D}$}}
\def\capE{\mbox{\boldmath${\mathsf E}$}}
\def\capF{\mbox{\boldmath${\mathsf F}$}}
\def\capG{\mbox{\boldmath${\mathsf G}$}}
\def\capH{\mbox{\boldmath${\mathsf H}$}}
\def\capI{\mbox{\boldmath${\mathsf I}$}}
\def\capK{\mbox{\boldmath${\mathsf K}$}}
\def\capL{\mbox{\boldmath${\mathsf L}$}}
\def\capM{\mbox{\boldmath${\mathsf M}$}}
\def\capR{\mbox{\boldmath${\mathsf R}$}}
\def\capW{\mbox{\boldmath${\mathsf W}$}}

\def\C{\mbox{\boldmath${\mathcal C}$}}
\def\E{\mbox{\boldmath${\mathcal E}$}}

\def\mA{\mbox{${\mathcal A}$}}
\def\mB{\mbox{${\mathcal B}$}}
\def\mC{\mbox{${\mathcal C}$}}
\def\mD{\mbox{${\mathcal D}$}}
\def\mE{\mbox{${\mathcal E}$}}
\def\mF{\mbox{${\mathcal F}$}}
\def\mG{\mbox{${\mathcal G}$}}
\def\mH{\mbox{${\mathcal H}$}}
\def\mI{\mbox{${\mathcal I}$}}
\def\mJ{\mbox{${\mathcal J}$}}
\def\mK{\mbox{${\mathcal K}$}}
\def\mL{\mbox{${\mathcal L}$}}
\def\mM{\mbox{${\mathcal M}$}}
\def\mQ{\mbox{${\mathcal Q}$}}
\def\mR{\mbox{${\mathcal R}$}}
\def\mS{\mbox{${\mathcal S}$}}
\def\mT{\mbox{${\mathcal T}$}}
\def\mV{\mbox{${\mathcal V}$}}
\def\mY{\mbox{${\mathcal Y}$}}
\def\mZ{\mbox{${\mathcal Z}$}}

\def\AAM{{\it Adv. Appl. Mech. }}
\def\AMM{{\it Acta Metall. Mater. }}
\def\ARMA{{\it Arch. Rat. Mech. Analysis }}
\def\AMR{{\it Appl. Mech. Rev. }}
\def\CMAME {{\it Comput. Meth. Appl. Mech. Engrg. }}
\def\CMT {{\it Cont. Mech. and Therm.}}
\def\CRAS{{\it C. R. Acad. Sci., Paris }}
\def\EFM{{\it Eng. Fract. Mech. }}
\def\EJMA{{\it Eur.~J.~Mech.-A/Solids }}
\def\IMA{{\it IMA J. Appl. Math. }}
\def\IJES{{\it Int. J. Engng. Sci. }}
\def\IJMS{{\it Int. J. Mech. Sci. }}
\def\IJNME{{\it Int. J. Numer. Meth. Eng. }}
\def\IJNAMG{{\it Int. J. Numer. Anal. Meth. Geomech. }}
\def\IJP{{\it Int. J. Plasticity }}
\def\IJSS{{\it Int. J. Solids Struct. }}
\def\IngA{{\it {Ing. Archiv }}}
\def\JACS{{\it J. Am. Ceram. Soc. }}
\def\JAM{{\it J. Appl. Mech. }}
\def\JAP{{\it J. Appl. Phys. }}
\def\JE{{\it J. Elasticity }}
\def\JM{{\it J. de M\'ecanique }}
\def\JMPS{{\it J. Mech. Phys. Solids. }}
\def\MOM{{\it Mech. Materials }}
\def\MRC{{\it Mech. Res. Comm. }}
\def\MSE{{\it Mater. Sci. Eng. }}
\def\MMS{{\it Math. Mech. Solids }}
\def\MPCPS{{\it Math. Proc. Camb. Phil. Soc. }}
\def\PRSA{{\it Proc. R. Soc. Lond., Ser. A}}
\def\PRSL{{\it Proc. R. Soc. Lond. }}
\def\QAM{{\it Quart. Appl. Math. }}
\def\QJMAM{{\it Quart. J. Mech. Appl. Math. }}
\def\ZAMP{{\it Z. angew. Math. Phys. }}


\def\salto#1#2{%
[\mbox{\hspace{-#1em}}[#2]\mbox{\hspace{-#1em}}]}


\title{\bf Anisotropic effective higher-order response
of heterogeneous Cauchy elastic  materials}
\author{M. Bacca, F. Dal Corso, D. Veber, D. Bigoni\footnote{Corresponding author} \\
Department of Civil, Environmental and Mechanical Engineering\\
University of Trento, \\ via Mesiano 77, I-38123 Trento, Italy\\
e-mail: mattia.bacca@ing.unitn.it, francesco.dalcorso@unitn.it, \\
daniele.veber@unitn.it, bigoni@unitn.it}
\maketitle

\begin{abstract}

The homogenization results obtained by Bacca et al. (Homogenization of heterogeneous Cauchy-elastic materials
leads to Mindlin second-gradient elasticity. Part I: Closed form expression for the effective higher-order
constitutive tensor. http://arxiv.org/abs/1305.2365 {\it Submitted}, 2013), to define effective
 second-gradient elastic materials from heterogeneous Cauchy elastic solids, are
extended here to the case of phases having non-isotropic tensors of inertia.
It is shown that the nonlocal constitutive tensor for the homogenized material depends on both
the inertia properties of the RVE and the difference between the effective and the matrix local elastic tensors.
Results show that: (i.) a composite material can be designed to result locally isotropic but
nonlocally orthotropic; (ii.) orthotropic nonlocal effects are introduced when a dilute distribution of aligned elliptical holes
and, in the limit case, of cracks is homogenized.

\end{abstract}


\section{Introduction}

Microstructures introduce length-scales and nonlocal effects in the mechanical modeling of solids,
so that the use of higher-order theories becomes imperative, in particular when large strain gradients are involved
[as in the case of localization of deformation (Dal Corso and Willis, 2011) or fracture mechanics (Mishuris et al., 2012)].

\begin{figure*}[!htcb]
  \begin{center}
\includegraphics[width=12 cm]{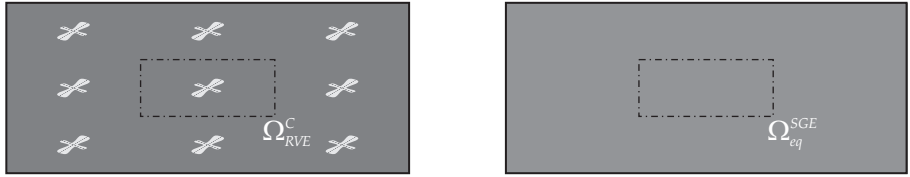}
\caption{\footnotesize Heterogeneous Cauchy material (left) made up of an inclusion phase with constitutive tensor $\capC^{(2)}$
embedded within a matrix with constitutive tensor $\capC^{(1)}$ and
homogeneous equivalent SGE material (right) with effective tensors $\capC^{eq}$ and $\capA^{eq}$.}
 \label{eteroomo_generic}
 \end{center}
\end{figure*}

\begin{figure*}[!htcb]
  \begin{center}
\includegraphics[width=13 cm]{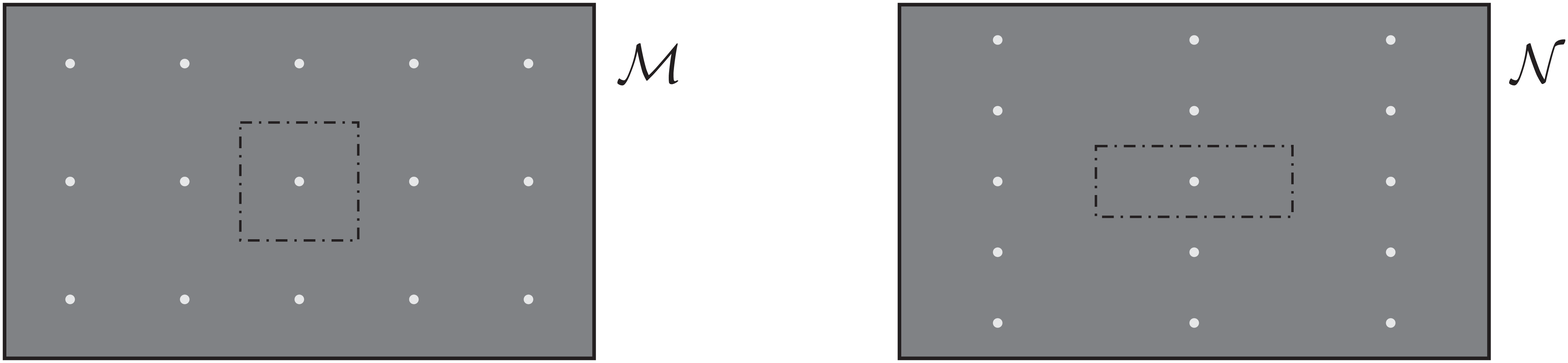}
\caption{\footnotesize Two composites $\mathcal{M}$ and $\mathcal{N}$
containing a dilute suspension of isotropic circular inclusions in an isotropic matrix (with the same volume ratio $f\ll1$).
While standard homogenization leads to the same effective isotropic constitutive fourth-order tensor for both distributions,
$\capC^{eq}(\mathcal{M})=\capC^{eq}(\mathcal{N})$,
second-order homogenization leads to different effective nonlocal properties, $\capA^{eq}(\mathcal{M})\neq\capA^{eq}(\mathcal{N})$.
In particular, while $\capA^{eq}(\mathcal{M})$ is isotropic, $\capA^{eq}(\mathcal{N})$ is orthotropic.}
 \label{tisdenti}
 \end{center}
\end{figure*}

Nonlocality, often introduced phenomenologically (Cosserat and Cosserat, 1909;
Koiter, 1964; Mindlin, 1964; Aifantis, 1978), can be analyzed from the
fundamental and scarcely explored\footnote{
The attempt of relating the microstructure to higher-order effect goes back to Dean and Urgate (1968) and
Banks and Sokolowski (1968) and has been pursued, among others, by (Berglund, 1982;
Pideri and Seppecher, 1997; Forest, 1998; Wang and Stronge, 1999; Ostoja-Starzewski et al., 1999;
Bouyge et al., 2001).}
point of view in which it is linked to the microstructure of a Cauchy elastic
heterogeneous material via homogenization theory. This approach was followed by Bigoni and Drugan (2007)
and extended by Bacca et al. (2013a) who obtained a closed-form solution for the nonlocal effective response through a
second-order homogenization procedure, based on the dilute approximation. In the present article
this result is generalized  to allow the possibility of phases having non-spherical ellipsoids   of inertia
(or non-circular ellipses of inertia in plane strain), Fig. \ref{eteroomo_generic}, left.
 By matching the elastic energy of a Cauchy elastic heterogeneous material (Fig. \ref{eteroomo_generic}, left) with that
 of a homogeneous higher-order elastic solid (Fig. \ref{eteroomo_generic}, right), both subject to the same displacement conditions on
 the boundary, it is shown that a sixth-order constitutive tensor can be obtained, defining a
 second-gradient elastic (SGE) material (Mindlin, 1964).
The results show how the shape and the constitutive symmetry class of the phases influence the higher-order response
and are exploited to investigate two particular cases of special interest.
One case is that a Cauchy elastic heterogeneous material (e.g. a rectangular lattice of circular elastic inclusions in
an isotropic elastic matrix, Fig. \ref{tisdenti}, right) can be designed to result at the same time locally isotropic, but nonlocally orthotropic.
The other one is the case of a dilute distribution of aligned elliptical voids and the limit case of cracks,
providing nonlocal orthotropic effects.

\section{Nonlocal effective response}

Through application of the second-order homogenization procedure, the evaluation
of the effective nonlocal response of a Second Gradient Elastic solid (Mindlin, 1964) obtained by Bacca et al. (2013a),
and valid for a RVE having inclusion and matrix with spherical ellipsoids of inertia,
is extended to the case of phases with generic ellipsoid of inertia.

The homogenization is performed on a heterogeneous Cauchy RVE (volume $\Omega_{RVE}^C$; Fig. \ref{eteroomo_generic}, left),
where a {\it dilute} distribution of  inclusions
(phase (2), local constitutive tensor $\capC^{(2)}$, volume $\Omega_{2}^C$=$f \Omega_{RVE}^C$, volume fraction $f\ll1$) is embedded within a matrix
(phase (1), local constitutive tensor $\capC^{(1)}$ volume $\Omega_{1}^C$=$(1-f) \Omega_{RVE}^C$).
The homogenization leads to a homogeneous SGE material
(with effective local constitutive tensor $\capC^{eq}$
and nonlocal constitutive tensor $\capA^{eq}$, volume $\Omega_{eq}^{SGE}=\Omega_{RVE}^C$; Fig. \ref{eteroomo_generic}, right).
The equivalent response is obtained through the annihilation of the strain energy mismatch between the heterogeneous and equivalent materials, which is
\beq
\lb{gapeqn}
\mathcal{W}_{RVE}^{C}\left(\capC^{(1)},\capC^{(2)}\right)-
\mathcal{W}_{eq}^{SGE}\left(\capC^{eq},\capA^{eq}\right)=0,
\eeq
when the following second-order (linear and quadratic) displacement field $\overline{\bu}$ is imposed on both the boundaries of the RVE and the SGE
\beq
\lb{boundaryconditions_aa}
\bu=\overline{\bu}, ~~~ \mbox{on } \partial\Omega_{RVE}^C, \qquad \mbox{and}
\qquad
\left\{
\begin{array}{lll}
\bu=\overline{\bu}, \\[4mm]
D \bu=D\overline{\bu},
\end{array}
\right.
~~~ \mbox{on } \partial\Omega_{eq}^{SGE},
\eeq
with
\beq\lb{displacements}
\overline{u}_i=\underbrace{\alpha_{ij} x_j}_{\ds \overline{u}_i^\alpha}+\underbrace{\beta_{ijk} x_j x_k}_{\ds \overline{u}_i^\beta},
\eeq
where $\alpha_{ij}$ and $\beta_{ijk}$ are constant coefficients, the latter satisfying the
symmetry $\beta_{ijk}$=$\beta_{ikj}$. The strain energies for the heterogeneous Cauchy material
and the equivalent SGE material are given by integration over the domain of
the respective strain energy density, namely
\beq
\lb{defenergysgm}
\begin{array}{lll}
\mathcal{W}_{RVE}^C=\ds\frac{1}{2}\int_{\Omega_{1}^C}\capC^{(1)}_{ijhk}u_{i,j}u_{h,k}
+\frac{1}{2}\int_{\Omega_{2}^C}\capC^{(2)}_{ijhk}u_{i,j}u_{h,k},\\[5mm]
\mathcal{W}_{eq}^{SGE}=\ds\frac{1}{2}\int_{\Omega_{eq}^{SGE}}\left[\capC^{eq}_{ijhk}u_{i,j}u_{h,k}
+\capA^{eq}_{ijklmn}u_{k,ij}u_{n,lm}\right].
\end{array}
\eeq

To explicitly evaluate the effective nonlocal response, some of the geometrical assumptions introduced by Bacca et al. (2013a)
are now modified.
In particular, coincidence of the centroids of matrix and inclusion,
both centered at the origin of the $x_i$--axes (i.e. geometrical property GP1 in
Bacca et al. 2013a) is still assumed, namely, the static moment vectors of the inclusion, of the matrix and of the RVE are null
\beq \lb{Eulerphases}
\bS(\Omega_{1}^C)=
\bS(\Omega_{2}^C)=
\bS(\Omega_{RVE}^C)=\b0, \qquad \bS(V)=\int_{V} \bx,
\eeq
but no restriction is now assumed about the shape and the directions of the ellipsoids of
inertia for matrix and the inclusion (i.e. the geometrical property GP2 of Bacca et al. 2013a is removed).

Introducing the normalized inertia  tensor $\bB$ for a generic solid occupying the region $V$, defined as
the second-order Euler tensor of inertia $\bE$ divided by the RVE volume (or area in plane-strain) $\Omega_{RVE}^C$,
\beq\label{euler}
\bB(V) =\frac{\bE(V)}{\Omega_{RVE}^C},\qquad \bE(V) =\int_{V} \bx\,\otimes\, \bx,
\eeq
the normalized tensors of inertia for the matrix $\bB^{(1)}=\bB(\Omega_{1}^C)$, the inclusion $\bB^{(2)}=\bB(\Omega_{2}^C)$
and the RVE $\bB^{RVE}=\bB(\Omega_{RVE}^C)$ are given by
\beq\label{inertianormalizedphases}
\begin{array}{lll}
\ds\bB^{(1)}=(1-f) \sum_{k=1}^{N}\left[\rho_{k}^{(1)}\right]^2 \,\be_{[k]}^{(1)} \otimes \be_{[k]}^{(1)},\\
\ds\bB^{(2)}=f \sum_{k=1}^{N}\left[\rho_{k}^{(2)}\right]^2 \,\be_{[k]}^{(2)} \otimes \be_{[k]}^{(2)},\\
\ds\bB^{RVE} =\sum_{k=1}^{N}\left[\rho_k^{RVE}\right]^2 \,\be_{[k]}^{RVE} \otimes \be_{[k]}^{RVE},
\end{array}
\eeq
where $\left\{\be_{[k]}^{(1)},\be_{[k]}^{(2)},\be_{[k]}^{RVE} \right\}$ are the $k$-th unit vectors identifying
the principal system for the matrix, the inclusion, and the RVE,
and $\left\{\rho_{k}^{(1)},\rho_{k}^{(2)},\rho_{k}^{RVE} \right\}$ are the respective radii of gyration,
and $N$  is equal to 3 (or 2 for plane strain case).

Note that from eqns (\ref{euler}) and (\ref{inertianormalizedphases}) it follows
\beq\label{sommadiB}
\bB^{RVE} = \bB^{(1)} + \bB^{(2)},
\eeq
so that the normalized tensors of inertia for the RVE can be isotropic, $\bB^{RVE} = \left(\rho^{RVE}\right)^2 \Id$,
 even when the tensors of inertia of the matrix and inclusion are not isotropic. In other words, since $\bB^{(1)}$, $\bB^{(2)}$ and
$\bB^{RVE}$ are geometrical quantities, $\bB^{RVE}$ corresponds to the normalized inertia of the region enclosed by the the external contour
of the RVE.

Considering the normalized inertia tensors (\ref{inertianormalizedphases}),
it is assumed that for the inclusion all the radii of gyration $\rho_k^{(2)}$ vanish in the limit of null inclusion volume fraction $f$
(a modification of the geometrical property GP3 in Bacca et al. 2013a), namely,
\beq\label{gp3}
\lim_{f\rightarrow0}\rho_k^{(2)}(f)=0,\qquad \forall \,\, k=1,...,N.
\eeq

Similarly to the computations in (Bacca et al., 2013a), the energy mismatch condition (\ref{gapeqn}) is now used to obtain
a nonlocal material equivalent to the heterogeneous Cauchy RVE.
Since Lemmas 1 and 2 in (Bacca et al., 2013a) are not affected by the geometrical
assumption GP2 (which is now removed),
the mutual energy is still null at first-order in $f$. Considering that $\capC^{eq}$ is obtained through a first-order
homogenization, the energy mismatch (\ref{gapeqn}) is given only by the contribution related to the sole
quadratic displacement boundary condition,
\beq
\lb{gapeqnsolobeta}
\mathcal{W}_{RVE}^{C}(\overline{\bu})-
\mathcal{W}_{eq}^{SGE}(\overline{\bu},D\overline{\bu})=
\mathcal{W}_{RVE}^{C}(\overline{\bu}^{\beta^{\diamond *}})-
\mathcal{W}_{eq}^{SGE}(\overline{\bu}^{\beta^{\diamond *}},D\overline{\bu}^{\beta^{\diamond *}})+o(f),
\eeq
where the symbol $\diamond$ denotes the fact that the quadratic displacement field is restricted to produce a self-equilibrated stress in
an auxiliary homogeneous medium with local tensor $\capC^*$, taken
as a first-order perturbation in $f$ to the equivalent local constitutive tensor $\capC^{eq}$ (see Bacca et al., 2013a, for details).

Lemmas 3 and 4 in (Bacca et al., 2013a) can be extended by
considering the definition of normalized Euler tensors of inertia given by eqns (\ref{inertianormalizedphases}), and from which
the strain energies instrumental to compute the energy mismatch can be obtained as
\beq
\lb{uno}
\mathcal{W}_{RVE}^{C}(\overline{\bu}^{\beta^{\diamond *}})=
2 \Omega_{RVE}^C B_{lm}^{RVE}\capC^{(1)}_{ijhk}\beta^{\diamond *}_{ijl}\beta^{\diamond *}_{hkm}+o(f),
\eeq
and
\beq
\lb{due}
\mathcal{W}_{eq}^{SGE}(\overline{\bu}^{\beta^{\diamond *}},D\overline{\bu}^{\beta^{\diamond *}})=
2 \Omega_{RVE}^C \left(B_{lm}^{RVE}\capC^{eq}_{ijhk}+\capA^{eq}_{jlikmh}\right)\beta^{\diamond
*}_{ijl}\beta^{\diamond *}_{hkm}+o(f).
\eeq

By using eqns (\ref{uno}) and (\ref{due}) to impose the annihilation of the energy mismatch (\ref{gapeqn})
for a generic purely quadratic boundary condition (and leading to a self-equilibrated stress within the arbitrary medium characterized by local tensor
$\capC^*$), the nonlocal sixth-order tensor $\capA^{eq}$ of the equivalent SGE material is evaluated (at first-order in $f$) as
\beq
\lb{sol}
\boxed{\begin{split}
\capA^{eq}_{ijhlmn}&=- \frac{f}{4}\left(
\tilde{\capC}_{ihln}B_{jm}^{RVE}+
\tilde{\capC}_{ihmn}B_{jl}^{RVE}+
\tilde{\capC}_{jhln}B_{im}^{RVE}+
\tilde{\capC}_{jhmn}B_{il}^{RVE}
\right),\end{split}}
\eeq
where $\tilde{\capC}$ is the first-order discrepancy tensor, defined as
\beq\label{valtari}
\capC^{eq}=\capC^{(1)}+f\tilde{\capC},
\eeq
and assumed to be known from standard homogenization.

Solution (\ref{sol}) represents the evaluation --in the dilute case-- of the nonlocal behaviour of the equivalent SGE material
under the geometrical assumptions (\ref{Eulerphases}) and (\ref{gp3}).
Regardless of the ellipsoid of inertia of the inclusion phase, this solution reduces to that given in (Bacca et al., 2013a) when a RVE
enclosing a region with a spherical ellipsoid of inertia, $B_{ij}^{RVE}=\left(\rho^{RVE}\right)^2 \delta_{ij}$, is considered\footnote{Note that in (Bacca et al., 2013a) the
spherical radius of gyration of the RVE is indicated with $\rho$, while within this article is denoted by $\rho^{RVE}$.}
\beq
\lb{solvecchia}
\begin{split}
\capA^{eq}_{ijhlmn}&=-f \frac{\left(\rho^{RVE}\right)^2}{4}\left(
\tilde{\capC}_{ihln}\delta_{jm}+
\tilde{\capC}_{ihmn}\delta_{jl}+
\tilde{\capC}_{jhln}\delta_{im}+
\tilde{\capC}_{jhmn}\delta_{il}
\right).
\end{split}
\eeq

It is worth to remark that, while the normalized inertia tensor $\bB^{RVE}$ is dependent only
on the shape and size of the external boundary of the RVE, the first-order discrepancy tensor $\tilde{\capC}$ is dependent on
the constitutive tensors of both phases and on the shape of the inclusion phase.

Similarly to the case of phases with spherical ellipsoid of inertia (Bacca et al., 2013b), it can be noted that:
\begin{itemize}
\item the equivalent SGE material is positive definite if and only if $\tilde{\capC}$ is negative definite;
\item the constitutive higher-order tensor $\capA^{eq}$ is linear in $f$ for dilute concentrations;
\end{itemize}
but, differently,
the higher-order material symmetries of the equivalent SGE solid involves {\it not only}
the material symmetries of the first-order discrepancy tensor $\tilde{\capC}$.
Indeed, when a RVE encloses a region with a non-spherical ellipsoid of inertia (see Appendix \ref{simmetrie} for details),
\begin{quote}
{\it
the symmetry class of the higher-order response
for the effective material coincides with the symmetry class of both the RVE's normalized inertia tensor $\bB^{RVE}$ and
the first-order discrepancy tensor $\tilde{\capC}$.
}
\end{quote}

Therefore it is transparent that an isotropic higher-order behaviour (in the dilute case) is obtained only when
both the first-order discrepancy tensor $\tilde{\capC}$ and the normalized inertia tensor $\bB^{RVE}$ are isotropic.

\section{Application cases}\lb{AppCases}

With reference to composites with phases having isotropic Cauchy behaviour
\beq
\lb{Isophases}
\ds\capC^{(1)}_{ijhk}\ds=\lambda_1 \delta_{ij} \delta_{hk}+\mu_1 (\delta_{ih}\delta_{jk}+\delta_{ik}\delta_{jh}), \qquad
\ds\capC^{(2)}_{ijhk}\ds=\lambda_2 \delta_{ij} \delta_{hk}+\mu_2 (\delta_{ih}\delta_{jk}+\delta_{ik}\delta_{jh}),
\eeq
the anisotropic nonlocal effective response is analyzed by considering applications
of the homogenization result (\ref{sol}) to the following specific geometries:
\begin{itemize}
\item RVE enclosing a region with a spherical ellipsoid of inertia;
\item inclusion  with non-spherical ellipsoid of inertia.
\end{itemize}
The above examples show that, similarly to the usual first-order homogenization, specific assumptions on the geometry of the phases lead to anisotropic
nonlocal effective response and that, more interestingly, in the case of RVE enclosing a region
 with non-spherical ellipsoid inertia (and under the dilute assumption),
anisotropic nonlocal effective response can arise even in composites having isotropic local effective behaviour.

\subsection{RVE enclosing a region with non-spherical ellipsoid inertia}
Restricting attention to inclusions which geometry is such that the
first-order discrepancy tensor, eqn (\ref{valtari}), is isotropic
\beq
\lb{CtildeIso}
\ds\tilde{\capC}_{ijhk}\ds=\tilde{\lambda} \delta_{ij} \delta_{hk}+\tilde{\mu} (\delta_{ih}\delta_{jk}+\delta_{ik}\delta_{jh}),
\eeq
and considering the principal directions of the RVE ellipsoid of inertia\footnote{
The nomenclature \lq RVE ellipsoid of inertia' means the inertia of the region enclosed within the external contour of the RVE,
according to eqn (\ref{sommadiB}).} as the reference system,
the components of the normalized tensor of inertia $\bB^{RVE}$,
eqn (\ref{inertianormalizedphases})$_3$, can be written as
\beq
B^{RVE}_{ij}=\sum_{k=1}^{N}\left(\rho^{RVE}_k\right)^2 \delta_{ik}\delta_{jk},
\eeq
and the solution (\ref{sol}) leads to the following effective  nonlocal orthotropic
tensor\footnote{The nonlocal parameters appearing in definition of the orthotropic tensor, eqn (\ref{solOrthGeom}),
have been denoted by  $\left\{a_2^{[k]};a_4^{[k]};a_5^{[k]}\right\}$ to use the same nomenclature as in (Bacca et al., 2013b).}
\beq\lb{solOrthGeom}
\begin{array}{lll}
\capA^{eq}_{ijhlmn}=\ds\sum_{k=1}^{N}&\ds\left\{
\frac{a_{2}^{[k]}}{2}\left(\delta_{ih}\delta_{kj}+\delta_{jh}\delta_{ki}\right)
\left(\delta_{ln}\delta_{km}+\delta_{mn}\delta_{kl}\right)
\right.\\[3mm]&\ds\left.+\frac{a_{4}^{[k]}}{2}\delta_{hn}\left[\delta_{kj}\left(\delta_{im}\delta_{kl}+\delta_{il}\delta_{km}\right)+\delta_{ki}\left(\delta_{jl}\delta_{km}+\delta_{jm}\delta_{kl}\right)\right]
\right.\\[3mm]&\ds\left.+\frac{a_{5}^{[k]}}{2}\left[\delta_{in}\delta_{kj}\left(\delta_{hl}\delta_{km}+\delta_{hm}\delta_{kl}\right)+\delta_{jn}\delta_{ki}\left(\delta_{hm}\delta_{kl}+\delta_{hl}\delta_{km}\right)\right]\right\},
\end{array}
\eeq
where the $3N$ nonlocal parameters are given by
\beq\lb{solOrthGeom1}
a_2^{[k]}=-f\frac{\left(\rho^{RVE}_k\right)^2}{2}\tilde{\lambda},\qquad a_4^{[k]}= a_5^{[k]}=-f\frac{\left(\rho^{RVE}_k\right)^2}{2}\tilde{\mu},
\qquad k=1,...,N.
\eeq
Note that the effective nonlocal tensor $\capA^{eq}$, eqn (\ref{solOrthGeom}), is an orthotropic sixth-order tensor
 and that it reduces to an isotropic  sixth-order tensor when the ellipsoid of inertia of the RVE becomes a sphere
 (or a circle in plane strain), $\rho_k^{RVE}=\rho^{RVE}$ for $k=1,..., N$. Furthermore, the
 homogenization procedure leads to a positive definite SGE material when
the discrepancy tensor $\tilde{\capC}$ is negative definite, corresponding to
\beq\begin{array}{c}\lb{posdefisoSol}
\tilde{K}<0,~~~
\tilde{\mu}<0,
\end{array}\eeq
where $\tilde{K}$ is the bulk modulus, equal to $\tilde{\lambda}+2\tilde{\mu}/3$ when $N=3$
and $\tilde{\lambda}+\tilde{\mu}$ when $N=2$.

It is worth to mention that the $3(N-1)$ nonlocal parameters
$\left\{a_{2}^{[k]};a_{4}^{[k]};a_{5}^{[k]}\right\}$ ($k=2,N$) can be expressed as functions of the two nonlocal parameters
$\left\{a_{2}^{[1]};a_{4}^{[1]}\right\}$ as
\beq
\lb{Sol2DOrthGeomciccia}
\begin{array}{lll}
 \ds a_{2}^{[k]}=\left(\frac{\rho^{RVE}_k}{\rho_1^{RVE}}\right)^2 a_{2}^{[1]},\qquad
 a_{4}^{[k]}=a_{5}^{[k]}=\left(\frac{\rho^{RVE}_k}{\rho_1^{RVE}}\right)^2 a_{4}^{[1]},\qquad\qquad k=2,N.
\end{array}
\eeq
Equation (\ref{Sol2DOrthGeomciccia}) reveals that (for inclusions less stiff than the matrix and) if $\rho^{RVE}_k/\rho_1^{RVE} > 1$ ($<1$), namely, when
the distance between the inclusions is larger (smaller) in the direction $k$ than in direction 1, then the nonlocal behaviour is stiffer (less stiff)
in the former direction than in the latter (Fig. \ref{tisdenti}, right).

The explicit evaluation of the non-local parameters is provided below for two simple cases,
by considering the discrepancy parameters $\tilde{\lambda}$ and $\tilde{\mu}$ reported in (Bacca et al., 2013b),
since the first-order homogenization result is not affected by the external shape of the RVE in the dilute case.

\paragraph{Circular elastic inclusion within a rectangular RVE.}
 In the case of a circular elastic inclusion (in plane strain condition) of radius $r$ embedded in
 a rectangular RVE with edges $h_1$ and $h_2$ (parallel to the directions $x_k$, $k=1,2$),
the  three nonlocal parameters
$\left\{a_{2}^{[1]};a_{4}^{[1]};a_{5}^{[1]}\right\}$
are obtained through eqn (\ref{solOrthGeom1}) as
\beq
\lb{Sol2DOrthGeom}
\begin{array}{lll}
 \ds a_{2}^{[1]}=\frac{\pi r^2}{24} \frac{h_1}{h_2}
\left[\frac{(K_1-K_2)(K_1+\mu_1)}{K_2+\mu_1}
-\frac{\mu_1(\mu_1-\mu_2)(K_1+\mu_1)}{2\mu_1\mu_2+K_1(\mu_1+\mu_2)}\right],\\[5mm]
 \ds a_{4}^{[1]}=a_5^{[1]}=\frac{\pi r^2}{24} \frac{h_1}{h_2}\frac{\mu_1(\mu_1-\mu_2)(K_1+\mu_1)}{2\mu_1\mu_2+K_1(\mu_1+\mu_2)},
 \end{array}
\eeq
from which, by using relation (\ref{Sol2DOrthGeomciccia}), the remaining three parameters  $\left\{a_{2}^{[2]};a_{4}^{[2]};a_{5}^{[2]}\right\}$
follow in the form
\beq
\ds a_{2}^{[2]}=\left(\frac{h_2}{h_1}\right)^2 a_{2}^{[1]}, \qquad a_{4}^{[2]}=a_{5}^{[2]}=\left(\frac{h_2}{h_1}\right)^2 a_{4}^{[1]}.
\eeq
Note that a positive definite equivalent SGE material is obtained only when the
inclusion phase is \lq softer' than the matrix in terms of \emph{both} shear and bulk moduli, namely,
\beq\label{mollorigido}
\mu_2<\mu_1,\qquad
K_2<K_1.
\eeq

\paragraph{Spherical inclusion within a parallelepipedic RVE.}
In the case of  a spherical elastic inclusion of radius $r$ embedded in a parallelepiped with sides $h_1$, $h_2$ and
$h_3$ (parallel to the directions $x_k$, $k=1,2,3$), the three nonlocal parameters
$\left\{a_{2}^{[1]};a_{4}^{[1]};a_{5}^{[1]}\right\}$
are obtained through eqn (\ref{solOrthGeom1}) as
\beq
\lb{Sol3DOrthGeom}
\begin{array}{lll}
\ds a_{2}^{[1]}=\frac{\pi r^3}{18} \frac{h_1}{h_2 h_3}\left[\frac{(3 K_1+4 \mu_1)(K_2-K_1)}{3K_2+4 \mu_1}
-\frac{2}{3}\frac{5\mu_1(\mu_2-\mu_1)(3 K_1+4 \mu_1)}{\mu_1 (3 K_1+4 \mu_2)+2 (3
K_1+4 \mu_1) (\mu_2+\mu_1)}\right],
\\[5mm]
\ds a_{4}^{[1]}=a_5^{[1]}=\frac{\pi r^3}{18} \frac{h_1}{h_2 h_3}\frac{5\mu_1(\mu_2-\mu_1)(3 K_1+4 \mu_1)}{\mu_1 (3 K_1+4 \mu_2)+2 (3 K_1+4 \mu_1) (\mu_2+\mu_1)},
\end{array}
\eeq
from which, by using relation (\ref{Sol2DOrthGeomciccia}), the remaining six
parameters $\left\{a_{2}^{[2]};a_{4}^{[2]};a_{5}^{[2]};a_{2}^{[3]};a_{4}^{[3]};a_{5}^{[3]}\right\}$
follow in the form
\beq
\begin{array}{lll}
\ds a_{2}^{[2]}=\left(\frac{h_2}{h_1}\right)^2 a_{2}^{[1]}, \qquad a_{4}^{[2]}=a_{5}^{[2]}=\left(\frac{h_2}{h_1}\right)^2 a_{4}^{[1]},
\qquad
\ds a_{2}^{[3]}=\left(\frac{h_3}{h_1}\right)^2 a_{2}^{[1]}, \qquad a_{4}^{[3]}=a_{5}^{[3]}=\left(\frac{h_2}{h_1}\right)^2 a_{4}^{[1]}.
\end{array}
\eeq
Similarly to the case of circular elastic inclusions, the equivalent SGE material is positive definite
 only when condition (\ref{mollorigido}) is satisfied.

\subsection{Inclusion  with non-spherical ellipsoid inertia}
It has been shown that solution (\ref{solvecchia}), obtained by Bacca et al. (2013a),
remains valid regardless of the ellipsoid of inertia of the inclusions,
whenever the RVE encloses a region with a spherical ellipsoid of inertia. Solution (\ref{solvecchia}) is now exploited to achieve the nonlocal effective behaviour
of a composite with aligned elliptical holes and, in the limit of vanishing ellipse axes ratio, of cracks.

\paragraph{Elliptical hole within an isotropic matrix.}
The presence of equally-oriented inclusions with non-spherical ellipsoid of inertia within a composite introduces an anisotropy in the effective behaviour.
In particular, when a system of dilute aligned elliptic holes
(with semi-axis $b_{k}$ parallel to direction $x_k$, $k=1,2$) is considered
within an isotropic matrix, the effective behaviour is orthotropic with orthotropy axes coincident with those of the elliptical hole,
so that the first-order discrepancy tensor $\tilde{\capC}$ described in the reference system of orthotropy is (Tsukrov and Kachanov, 2000)
\beq
\label{ortopedico}
\begin{array}{lll}
\ds\tilde{\capC}_{ijhk}\ds=&
\tilde{\lambda} \delta_{ij} \delta_{hk}+\tilde{\mu} (\delta_{ih}\delta_{jk}+\delta_{ik}\delta_{jh})
+\tilde{\xi}\left(\delta_{i1}\delta_{j2}+\delta_{i2}\delta_{j1}\right)
\left(\delta_{h1}\delta_{k2}+\delta_{h2}\delta_{k1}\right)
+\tilde{\omega}\delta_{i1}\delta_{j1}\delta_{h1}\delta_{k1},
\end{array}
\eeq
with
\beq\lb{discrepanzeortotropo}
\begin{array}{rl}
\tilde{\lambda}&\ds=-(\lambda_1 +2 \mu_1 )
\frac{ \lambda_1\left(\lambda_1+2\mu_1\right)\left(1+\Lambda^2\right)      -2 \Lambda \mu_1 ^2}{2 \Lambda \mu_1  (\lambda_1 +\mu_1 )},\\[4mm]
\tilde{\mu}&\ds=-\left(1+\Lambda\right)\left(\lambda_1 +2 \mu_1 \right)\frac{ \lambda_1(1-\Lambda) +2 \mu_1}{2 \Lambda \left(\lambda_1 +\mu_1\right)},\\[4mm]
\tilde{\xi}&\ds=\left(1-\Lambda^2\right)\frac{\lambda_1 +2 \mu_1}{2 \Lambda},\\[4mm]
\tilde{\omega} &\ds=\left(1-\Lambda^2\right) \frac{\lambda_1 +2 \mu_1}{\Lambda},
\end{array}\eeq
where the parameter $\Lambda$ is the ratio between the ellipse's semi-axes, $\Lambda=b_{2}/b_{1}$.

Considering that the RVE encloses a region with a spherical ellipsoid of inertia, $B_{ij}^{RVE}=\left(\rho^{RVE}\right)^2\delta_{ij}$,
the effective nonlocal tensor $\capA^{eq}$ can be obtained, exploiting solution (\ref{sol}), as
the following positive-definite orthotropic sixth-order
tensor\footnote{The five nonlocal parameters appearing in definition of the orthotropic tensor, eqn (\ref{aeqorthotropicsolution}),
have been denoted by $\left\{a_2;a_4;a_5;a_6;a_9\right\}$ to use the same nomenclature as in (Bacca et al., 2013b).}

\beq
\lb{aeqorthotropicsolution}
\begin{array}{llll}
\ds\capA^{eq}_{ijhlmn}
=&\ds\frac{a_2}{2}
\left[\delta_{ih}\left(\delta_{jl}\delta_{mn}+\delta_{jm}\delta_{ln}\right)
    +\delta_{jh}\left(\delta_{il}\delta_{mn}+\delta_{im}\delta_{ln}\right)\right]
\\[3mm]
&\ds+a_4
\left(\delta_{il}\delta_{jm}+
     \delta_{im}\delta_{jl}\right)\delta_{hn}+\frac{a_5}{2}
\left[\delta_{in}\left(\delta_{jl}\delta_{hm}+\delta_{jm}\delta_{hl}\right)
    +\delta_{jn}\left(\delta_{il}\delta_{hm}+\delta_{im}\delta_{hl}\right)\right]
\\[3mm]
&+\ds\frac{a_6}{2}
\left\{\left(\delta_{i1}\delta_{h2}+\delta_{i2}\delta_{h1}\right)\left[
\left(\delta_{l1}\delta_{n2}+\delta_{l2}\delta_{n1}\right)\delta_{jm}
+\left(\delta_{m1}\delta_{n2}+\delta_{m2}\delta_{n1}\right)\delta_{jl}
\right]\right.\\[3mm]&\ds\left.
+\left(\delta_{j1}\delta_{h2}+\delta_{j2}\delta_{h1}\right)\left[
\left(\delta_{l1}\delta_{n2}+\delta_{l2}\delta_{n1}\right)\delta_{im}
+\left(\delta_{m1}\delta_{n2}+\delta_{m2}\delta_{n1}\right)\delta_{il}
\right]\right\}
\\[3mm]
&+\ds\frac{a_9}{2}
\left[\delta_{i1}\left(\delta_{l1}\delta_{jm}+
\delta_{m1}\delta_{jl}\right)
+\delta_{j1}\left(\delta_{l1}\delta_{im}
+\delta_{m1}\delta_{il}\right)\right]\delta_{h1}\delta_{n1},
\end{array}
\eeq

with the following nonlocal parameters
\beq\begin{array}{ccc}\lb{nonlocalortotropo}
a_2=\ds-f\frac{\left(\rho^{RVE}\right)^2}{2}\tilde{\lambda},\,\,\,
a_4=a_5=\ds-f\frac{\left(\rho^{RVE}\right)^2}{2}\tilde{\mu}, \,\,\,
a_{6}=\ds-f\frac{\left(\rho^{RVE}\right)^2}{2}\tilde{\xi},\,\,\,
a_{9}=\ds-f\frac{\left(\rho^{RVE}\right)^2}{2}\tilde{\omega}.
\end{array}\eeq

In the case that the RVE has a square boundary, and by considering the first-order discrepancy
quantities given by eqn (\ref{discrepanzeortotropo}),
 the nonlocal parameters simplify to
\beq\begin{array}{lll}\lb{nonlocalortotropo2}
a_2=\ds\frac{\pi b_1^2}{48}
\frac{ \lambda_1\left(\lambda_1+2\mu_1\right)\left(1+\Lambda^2\right)-2 \Lambda \mu_1 ^2}{\mu_1  (\lambda_1 +\mu_1 )}
(\lambda_1 +2 \mu_1 ),\\[5mm]
a_4=a_5=\ds\frac{\pi b_1^2}{48}
\frac{ \lambda_1(1-\Lambda) +2 \mu_1}{\left(\lambda_1 +\mu_1\right)}\left(1+\Lambda\right)\left(\lambda_1 +2 \mu_1 \right), \\[5mm]
a_{6}=\ds-\frac{\pi b_1^2}{48}\left(1-\Lambda^2\right)\left(\lambda_1 +2 \mu_1\right),
\\[5mm]
a_{9}=\ds-\frac{\pi b_1^2}{24}\left(1-\Lambda^2\right)\left(\lambda_1 +2 \mu_1\right).
\end{array}\eeq

\begin{figure*}[!htcb]
  \begin{center}
\includegraphics[width=8 cm]{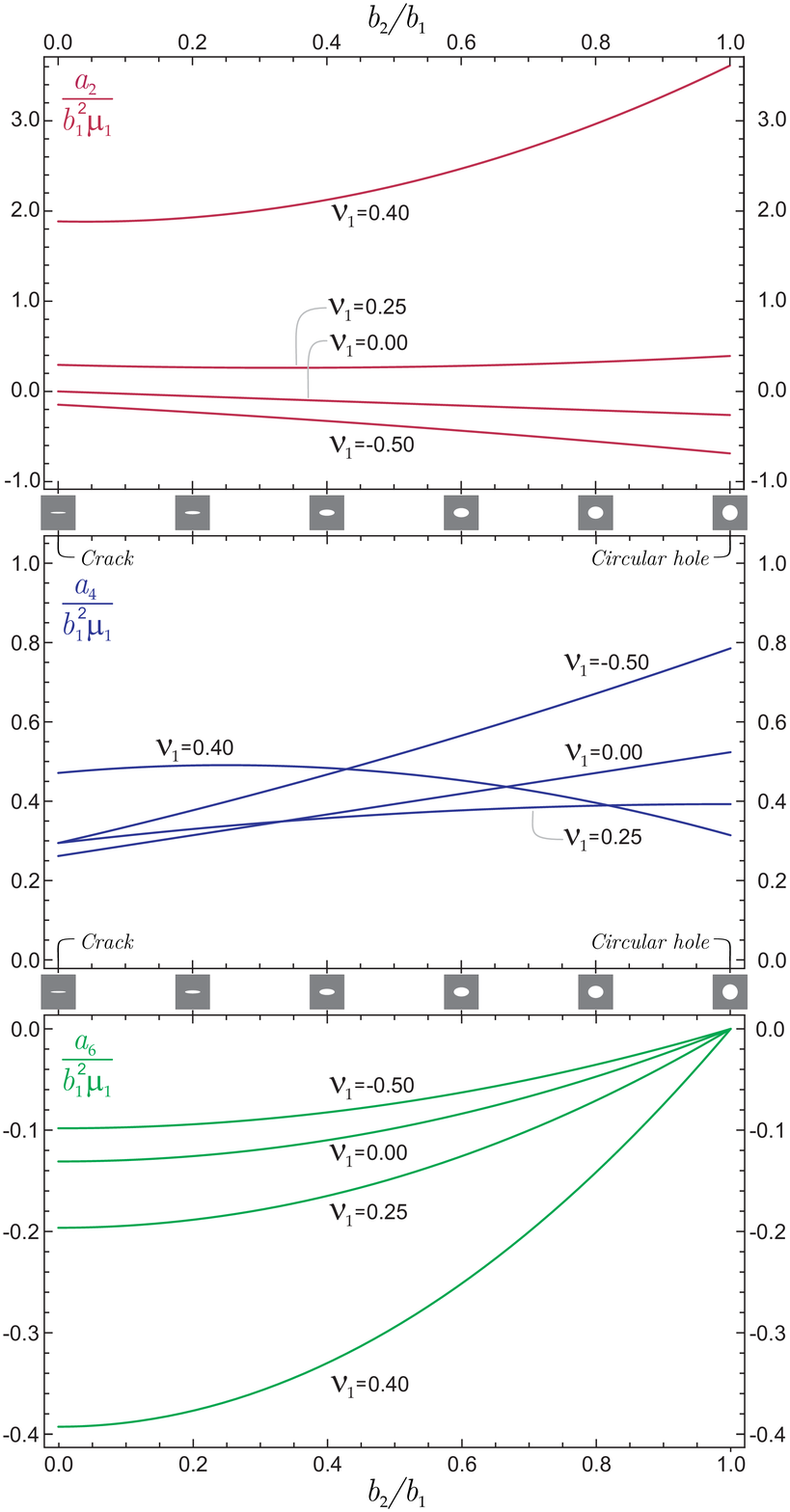}
\caption{\footnotesize Higher-order constitutive parameters $a_2$, $a_4=a_5$ and $a_6=a_9/2$ of the SGE solid equivalent
to a composite made up of an isotropic matrix containing a dilute suspension of aligned elliptical voids (with major semi-axis $b_1$),
as functions of the semi-axes ratio $b_2/b_1$, and for different values of the Poisson's ratio
of the matrix $\nu_1$ =$\left\{-0.5;0;0.25;0.4\right\}$. The constants are made dimensionless through
division by parameter $b_1^2\mu_1$. Note that both constants $a_6$ and $a_9$ approach zero in the limit  $b_2/b_1 \rightarrow 1$, so that the nonlocal
orthotropic behaviour turns out to become isotropic when the elliptical voids become circular holes.}
 \label{a2a4a6}
 \end{center}
\end{figure*}

The nonlocal effective constants $\left\{a_2;a_4=a_5;a_{6}=a_{9}/2\right\}$ given by eqn (\ref{nonlocalortotropo2})
are reported in Fig. \ref{a2a4a6}
as functions of the ellipse semi-axes ratio $b_2/b_1$ and for different values of the Poisson's ratio
of the matrix $\nu_1=\lambda_1/(2(\lambda_1+\mu_1))$.
Note that both the constants $a_6$ and $a_9$, defining the orthotropic nonlocal effective behaviour,
approach zero in the limit $b_2/b_1 \rightarrow 1$, so that the nonlocal
orthotropic behaviour turns out to be isotropic when the elliptical voids become circular holes.

\paragraph{Effective nonlocal parameters in the limit case of aligned cracks.}
Although the sixth-order tensor (\ref{aeqorthotropicsolution}) has been obtained excluding the possibility that
inclusions have non-null radius of inertia at null volume ratio (namely, eqn (\ref{gp3})),
the nonlocal  parameters of the effective tensor $\capA^{eq}$ for a dilute distribution of aligned ellipses, eqn (\ref{nonlocalortotropo2}),
can be used to obtain the limit values for a dilute distribution of aligned cracks ($\Lambda=b_2/b_1 \rightarrow 0$)
with length $2b_1$ within a square RVE as
\beq
\lb{nonlocalortotropocrack}
a_2=\ds\frac{\pi b_1^2}{48}
\frac{\lambda_1  (\lambda_1+2 \mu_1)^2}{ \mu_1  (\lambda_1 +\mu_1 )},\,\,\,
a_4=a_5=\ds\frac{\pi b_1^2}{48}
\frac{\left(\lambda_1 +2 \mu_1 \right)^2}{ \lambda_1 +\mu_1},\,\,\,
a_{6}=-\ds\frac{\pi b_1^2}{48}(\lambda_1 +2 \mu_1),\,\,\,
a_{9}=-\ds\frac{\pi b_1^2}{24}(\lambda_1 +2 \mu_1).
\eeq

\section{Conclusions}

An homogenization scheme has been shown to describe nonlocal effects for a heterogeneous Cauchy elastic material containing a {\it dilute}
distribution of inclusions of arbitrary shape in an arbitrary RVE (under the condition that inclusion and RVE have a coincident center of mass).
Results show that the shape of the RVE influences the effective nonlocal behaviour even under the dilute approximation, in contrast
with the usual homogenization scheme valid at first-order.

\vspace*{12mm} \noindent {\sl Acknowledgments}
M. Bacca gratefully acknowledges financial support from Italian Prin 2009 (prot. 2009XWLFKW-002).
D. Bigoni, F. Dal Corso and D. Veber  gratefully acknowledge financial support from the grant PIAP-GA-2011-286110-INTERCER2,
\lq Modelling and optimal design of ceramic structures with defects and imperfect interfaces'.
\vspace*{10mm}

 { \singlespace
}

\appendix

\section{Coincidence of symmetry class of $\capA^{eq}$ with the
intersection of the symmetry classes of $\tilde{\capC}$ and $\bB^{RVE}$}\label{simmetrie}

\setcounter{equation}{0}
\renewcommand{\theequation}{{A}.\arabic{equation}}

Considering the sixth-order tensor $\capD^{eq}$ defined as
\beq\label{deq}
\capD^{eq}_{ijhlmn}=-f \tilde{\capC}_{ihln}B^{RVE}_{jm},
\eeq
and having the following symmetries
\beq
\capD^{eq}_{ijhlmn}=\capD^{eq}_{lmnijh}=\capD^{eq}_{hjilmn}=\capD^{eq}_{ijhnml},
\eeq
the sixth-order tensor $\capA^{eq}$, eqn (\ref{sol}),
is obtained through application of the symmetrization operator $\mathcal{S}$ on  $\capD^{eq}$ as follows
\beq\label{simm12}
\capA^{eq}=\mathcal{S}(\capD^{eq}), \qquad
\capA^{eq}_{ijhlmn}=\ds\frac{1}{4}\left(\capD^{eq}_{ijhlmn}+\capD^{eq}_{ijhmln}+\capD^{eq}_{jihlmn}+\capD^{eq}_{jihmln}\right),
\eeq
so that $\capA^{eq}$ has the following symmetries
\beq
\capA^{eq}_{ijhlmn}=\capA^{eq}_{lmnijh}=\capA^{eq}_{jihlmn}=\capA^{eq}_{ijhmln}.
\eeq
The relation (\ref{simm12}) can be inverted through the inverse symmetrization operator $\mathcal{S}^{-1}$ defined as
\beq\label{simm13}
\begin{array}{lll}
\capD^{eq}=\mathcal{S}^{-1}(\capA^{eq}),\qquad
\capD^{eq}_{ijhlmn}=&
\capA^{eq}_{ijhlmn}
+\capA^{eq}_{jhimnl}
+\capA^{eq}_{hijnlm}
-\capA^{eq}_{ijhnlm}
-\capA^{eq}_{hijlmn}\\[3mm]
&+\capA^{eq}_{ijhmnl}
+\capA^{eq}_{jhilmn}
-\capA^{eq}_{jhinlm}
-\capA^{eq}_{hijmnl}.
\end{array}
\eeq

A material symmetry (with respect to an orthogonal transformation) for a tensor $\capM$  corresponds to the condition
\beq
\mathcal{Q}(\capM)=\capM,
\eeq
where the orthogonal transformation $\mathcal{Q}$ (defined with respect to an orthogonal tensor $\bQ$) applied
to the RVE's normalized inertia tensor $\bB^{RVE}$, to the first-order discrepancy tensor $\tilde{\capC}$, and to the sixth-order tensors $\capA^{eq}$
and $\capD^{eq}$ is given by
\beq
\begin{array}{lll}
&\left[\mathcal{Q}(\bB^{RVE})\right]_{ij}=Q_{ip}Q_{jq} \bB^{RVE}_{pq},\qquad
&\left[\mathcal{Q}(\tilde{\capC})\right]_{ijhl}=Q_{ip}Q_{jq}Q_{hr}Q_{ls} \tilde{\capC}_{pqrs},\\[3mm]
&\left[\mathcal{Q}(\capA^{eq})\right]_{ijhlmn}=Q_{ip}Q_{jq}Q_{hr}Q_{ls}Q_{mt}Q_{nu} \capA^{eq}_{pqrstu},\qquad
&\left[\mathcal{Q}(\capD^{eq})\right]_{ijhlmn}=Q_{ip}Q_{jq}Q_{hr}Q_{ls}Q_{mt}Q_{nu} \capD^{eq}_{pqrstu}.
\end{array}
\eeq

Since the symmetrization  and the inverse symmetrization operators,
$\mathcal{S}$ and $\mathcal{S}^{-1}$,
are commutative with the orthogonal operator $\mathcal{Q}$,
\beq
\mathcal{Q}\left(\mathcal{S}(\capD^{eq})\right)=
\mathcal{S}\left(\mathcal{Q}(\capD^{eq})\right), \qquad
\mathcal{Q}\left(\mathcal{S}^{-1}(\capA^{eq})\right)=
\mathcal{S}^{-1}\left(\mathcal{Q}(\capA^{eq})\right),
\eeq
the symmetry class of $\capA^{eq}$ is coincident to that of $\capD^{eq}$, namely
\beq
\mathcal{Q}(\capA^{eq})=\capA^{eq}
\qquad\Leftrightarrow\qquad
\mathcal{Q}(\capD^{eq})=\capD^{eq},
\eeq
which, considering that the sixth-order tensor $\capD^{eq}$ is given by eqn (\ref{deq}),
coincide with the intersection of symmetry classes of $\tilde{\capC}$ and $\bB^{RVE}$, namely
\beq
\mathcal{Q}(\capA^{eq})=\capA^{eq},
\qquad\Leftrightarrow\qquad
\mathcal{Q}(\tilde{\capC})=\tilde{\capC}
\qquad \mbox{and}\qquad
\mathcal{Q}(\bB^{RVE})=\bB^{RVE}.
\eeq


\begin{thebibliography}{}





\bibitem{aifantis} Aifantis, E.C. (1978) A proposal for continuum with microstructure
\emph{Mech. Res. Comm.,} \textbf{5} (3), 139--145.

\bibitem{bacca}  Bacca, M., Bigoni, D., Dal Corso, F. and Veber, D. (2013a)
Homogenization of heterogeneous Cauchy-elastic materials leads to
Mindlin second-gradient elasticity. Part I: Closed form expression
for the effective higher-order constitutive tensor. {\it Submitted},
http://arxiv.org/abs/1305.2365.

\bibitem{bacca}  Bacca, M., Bigoni, D., Dal Corso, F. and Veber, D. (2013b)
Homogenization of heterogeneous Cauchy-elastic materials
leads to Mindlin second-gradient elasticity.
Part II: Higher-order constitutive properties and application cases. {\it Submitted}, http://arxiv.org/abs/1305.2380.

\bibitem{Banks}  Banks, C.B., and Sokolowski, U. (1968) On Certain Two-Dimensional Applications
of Couple-Stress Theory, \IJSS, \textbf{4}, 15-–29.


\bibitem{Berglund}  Berglund, K. (1982) Structural Models of Micropolar Media,  \emph{Mechanics of
Micropolar Media} (CISM Lecture Notes), O. Brulin and R. K. T. Hsieh, eds.,
World Scientific, Singapore, pp. 35–-86.

\bibitem{Bigoni} Bigoni, D., and Drugan, W.J. (2007), Analytical derivation of Cosserat moduli via homogenization of heterogeneous
elastic materials. \JAM, \textbf{74}, 741--753.

\bibitem{bouyge}  Bouyge, F., Jasiuk, I., and Ostoja-Starzewski, M. (2001), A Micromechanically
Based Couple-Stress Model of an Elastic Two-Phase Composite, \IJSS, \textbf{38}, 1721-–1735.


\bibitem{cosserat} Cosserat, E., and Cosserat, F., 1909, {\it Sur la th\'{e}orie des corps d\'{e}formables}, Herman, Paris.

\bibitem{dalcorso} Dal Corso, F. and Willis, J.R. (2011), Stability of strain gradient plastic materials.
\JMPS, \textbf{59}, 1251--1267.

\bibitem{Dean}  Dean, D.L., and Urgate, C.P. (1968), Field Solutions for Two-Dimensional
Frameworks, \emph{Int. J. Mech. Sci.}, \textbf{10}, 315-–339.


\bibitem{forest} Forest, S. (1998), Mechanics of Generalized Continua: Construction by Homogenization, \emph{J. Phys. IV}, \textbf{8}, 39-–48.

\bibitem{koiter} Koiter, W.T. (1964), Couple-Stresses in the Theory of Elasticity, Parts I and
II. {\it Proc. K. Ned. Akad. Wet.}, Ser. B: Phys. Sci., 67, 17-–44.

\bibitem{Mindlin} Mindlin, R.D. (1964), Micro-structure in linear elasticity.
\emph{Archs ration. Mech. Analysis}, \textbf{16}, 51--78.

\bibitem{Mishuris} Mishuris, G., Piccolroaz, A., and Radi, E. (2012), Steady-state propagation
of a Mode III crack in couple stress elastic materials.
\emph{Int. J. Eng. Sci.}, \textbf{61}, 112--128.

\bibitem{Ostoja} Ostoja-Starzewski, M., Boccara, S., and Jasiuk, I. (1999), Couple-Stress
Moduli and Characteristic Length of Composite Materials, \MRC, \textbf{26}, 387–397.

\bibitem{pideri} Pideri, C., and Seppecher, P. (1997) A second gradient material resulting from the homogenization
of an heterogeneous linear elastic medium. \CMT \textbf{9}, 241-–257.

\bibitem{wang}  Wang, X.L., and Stronge, W.J. (1999), Micropolar Theory for Two-
Dimensional Stresses in Elastic Honeycomb, \PRSA, \textbf{445}, 2091-–2116.


\bibitem{Tsukrov} Tsukrov, I. and Kachanov, M.  (2000) Effective moduli of an anisotropic material with elliptical
holes of arbitrary orientational distribution. {\it Int. J. Sol. Struct.}, \textbf{37},  5919--5941.


\end{thebibliography}
\end{document}